\documentclass[conference]{IEEEtran}
\usepackage{graphicx}
\usepackage{cite}
\usepackage{mathtools}
\usepackage{blindtext}
\usepackage{amsmath,amssymb,amsfonts}
\usepackage{algorithm}
\usepackage{algpseudocode}
\usepackage{setspace}
\usepackage{subcaption}	
\usepackage{textcomp}
\usepackage{balance}
\usepackage{xcolor}
\usepackage{multirow}
\usepackage{stfloats}
\usepackage{amsmath}
\usepackage{array}
\usepackage{url}
\usepackage{soul}
\usepackage{tabularx}
\usepackage{bm}
\usepackage{pifont}%
\renewcommand{\baselinestretch}{0.97}
\def\BibTeX{{\rm B\kern.05em{\sc i\kern-.025em b}\kern-.08em T\kern-.1667em\lower.7ex\hbox{E}\kern-.125emX}}

\begin{document}

\title{Sub-Terahertz and mmWave Penetration Loss Measurements for Indoor Environments}

\author{
\IEEEauthorblockN{Kairui Du, Ozgur Ozdemir, Fatih Erden, and 
Ismail Guvenc}
\IEEEauthorblockA{Department of Electrical and Computer Engineering, North Carolina State University, Raleigh, NC}
Email: \{kdu, oozdemir, ferden, iguvenc\}@ncsu.edu
}

\maketitle

\begin{abstract}

Millimeter-wave~(mmWave) and terahertz~(THz) spectrum can support significantly higher data rates compared to lower frequency bands and hence are being actively considered for 5G wireless networks and beyond.
These bands have high free-space path loss (FSPL) in line-of-sight (LOS) propagation due to their shorter wavelength.
Moreover, in non-line-of-sight (NLOS) scenario, these two bands suffer higher penetration loss than lower frequency bands which could seriously affect the network coverage.
It is therefore critical to study the NLOS penetration loss introduced by different building materials at mmWave and THz bands, to help establish link budgets for an accurate performance analysis in indoor environments.
In this work, we measured the penetration loss and the attenuation of several common constructional materials at mmWave (28 and 39~GHz) and sub-THz (120 and 144~GHz) bands. Measurements were conducted using a channel sounder based on NI PXI platforms. Results show that the penetration loss changes extensively based on the frequency and the material properties, ranging from 0.401~dB for ceiling tile at 28~GHz, to 16.608~dB for plywood at 144~GHz. Ceiling tile has the lowest measured attenuation at 28~GHz, while clear glass has the highest attenuation of 27.633~dB/cm at 144~GHz. As expected, the penetration loss and attenuation increased with frequency for all the tested materials.

\end{abstract}


\section{Introduction}

In recent years, there has been a rapid increase in the number of wireless user devices and a surging demand of higher data throughput, making the sub-6 GHz frequency band more congested and incapable of supporting the future data rate demand. Researchers have thus been motivated to focus on higher frequencies in the millimeter-wave~(mmWave)  and terahertz~(THz) spectrum. These higher frequency bands range from 30~GHz to 10~THz, have large amount of available bandwidth compared to the sub-6~GHz bands, and hence they offer a promising solution to the increasing data rate demand. 
For example, IEEE Communications Society recently identified THz connectivity as one of the nine communication technology trends to follow ~\cite{THz}. 
However, as the communication frequency increases, the wavelength becomes smaller, which  
significantly increases the free-space path loss~(FSPL) and the penetration loss in mmWave band and THz bands,  both for line-of-sight~(LOS) and non-line-of-sight~(NLOS) propagation scenarios.

Propagation and attenuation over common mmWave bands have recently been studied extensively, following the opening of spectrum above 24~GHz by the~FCC~\cite{FCC_28G}. Reflection and penetration loss measurements at 28~GHz in New York urban environment~\cite{NYC} showed that the outdoor building materials are good reflectors with the largest measured reflection coefficient of 0.896 for tinted glass, while indoor materials were found to be less reflective. A high penetration loss was observed with the largest to be 45.1~dB for three interior walls of the office building at 11.39~m separation between the transmitter~(Tx) and receiver~(Rx). In~\cite{NYC2}, the authors found that steerable antennas could help generate rich number of resolvable multipath components at 28~GHz in both LOS and NLOS scenarios and thus creating viable links with link distances on the order of 200~m. Foliage and ground reflection measurements at 73~GHz~\cite{foliage} showed an average of 0.4~dB/m foliage attenuation, and reflection coefficients ranged from 0.02 to 0.34 for dirt and gravel ground. The authors in~\cite{khatun2019indoor} studied indoor and outdoor penetration loss at 73~GHz and 81~GHz and reported an average penetration loss of 2~to 9~dB for wood and glass. They also observed outdoor materials had larger penetration losses. In~\cite{hosseini2020attenuation}, authors focused on penetration loss of several typical building materials in three popular mmWave bands (i.e., 28, 73, 91 GHz). Higher penetration loss was observed as the communication frequency was increased.

Directional and omnidirectional path loss models are presented in~\cite{rappaport2015wideband}, which developed temporal and spatial channel models and outage probabilities based on a large number of wideband channel measurements at 28, 38, 60, and 73~GHz. Another study~\cite{0.5-100model} focused on different theoretical mmWave propagation channel models to compare LOS propagation, large-scale path loss, and building penetration loss  over the 0.5-100~GHz frequency range. In~\cite{directional}, empirical path loss models at 60 and 73 GHz were tuned using slope correction factors to accurately estimate mmWave path loss. 
In our recent work ~\cite{library}, 
we introduce models for power angular-delay profile (PADP) and large-scale path loss for both LOS and NLOS propagation, based on extensive indoor measurements at 28~GHz.
In~\cite{wahab_indoor,wahab_outdoor}, coverage enhancement for 28~GHz mmWave NLOS propagation using passive reflector was studied. It was found that the measured received power with reflectors could approach the received power of FSPL model at the same travel distance as the NLOS signal.

There are only a limited studies on wireless propagation and penetration at  sub-THz bands, which are expected to be used by the wireless technologies beyond 5G.
An overview in~\cite{opportunity&challenge} described many technical challenges and opportunities for wireless communication and sensing applications above 100~GHz and presented new propagation and penetration loss models. In~\cite{multti-ray_THz}, a multi-ray channel model considering the LOS, reflected, scattered, and diffracted paths in THz band was given and the THz channel characteristics were analyzed based on this model. Another study~\cite{propagation_model} provided deterministic and statistical channel models in the THz bands and presented hybrid methods, physical parameters of the THz channel and their implications for wireless communication design.

Few measurements at THz bands have been conducted so far. Authors in~\cite{xing2019indoor} focused on indoor reflection, penetration, scattering and path loss properties at both mmWave (28 and 73~GHz) and sub-THz (140~GHz) frequencies. The authors observed that reflection loss is lower at higher frequencies at a given incident angle while the penetration loss increases with frequency. Large-scale indoor path loss models at 140~GHz are provided and revealed a similar path loss exponent as observed at 28 and 73~GHz. Another study~\cite{0.1-2THz} on penetration loss reported that the loss at the lower end of the THz band is much lower than the loss at higher frequencies, and that  the incident angle affects the \emph{path length} inside the material. Undoubtedly, further measurement results are needed to develop better understanding and modeling of THz signal propagation. 

\begin{figure}
	\centering
	\begin{subfigure}{0.24\textwidth}
	\centering
	\includegraphics[width=\textwidth]{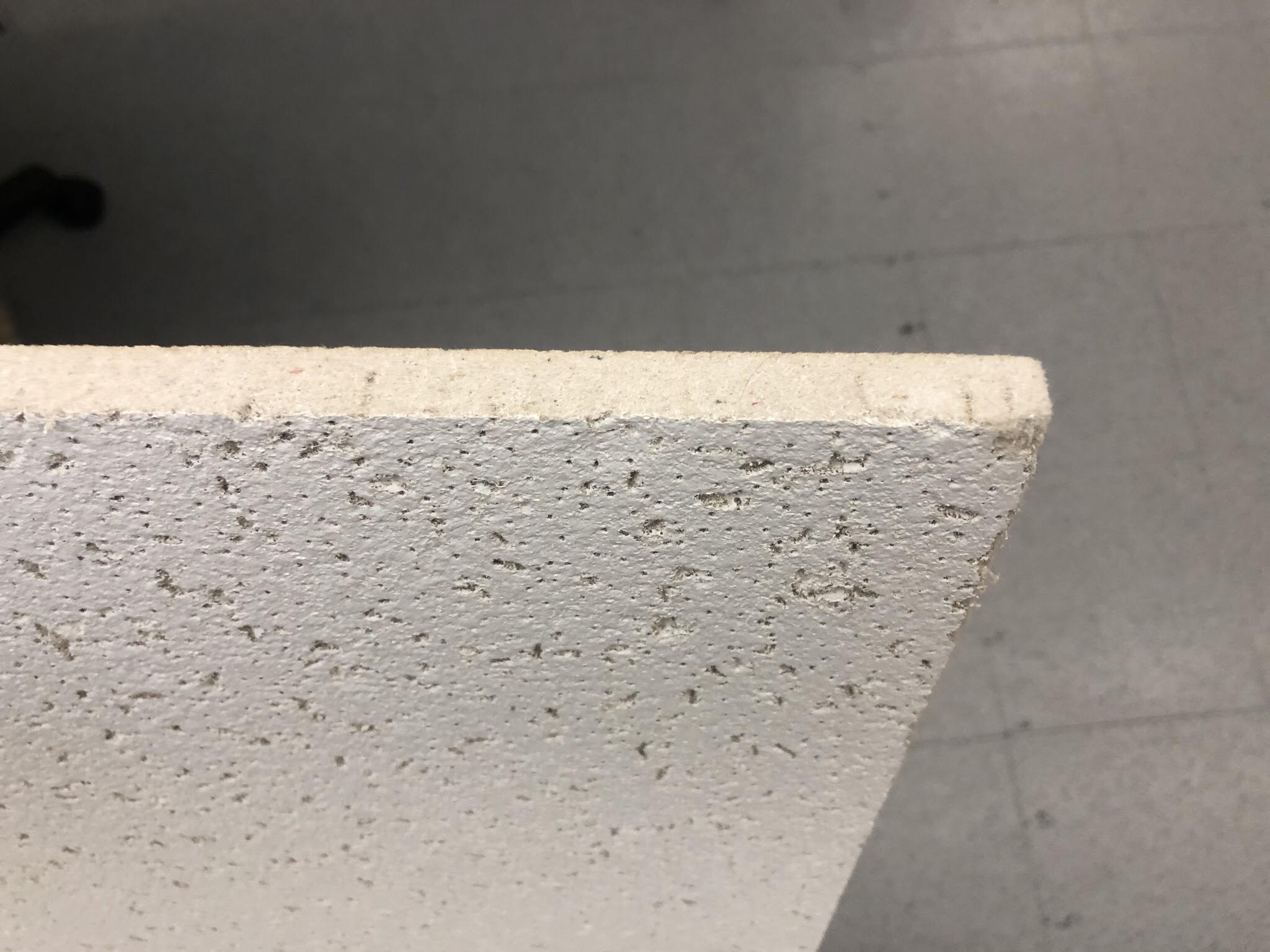}
	\caption{}
    \end{subfigure}
	\begin{subfigure}{0.24\textwidth}
	\centering
	\includegraphics[width=\textwidth]{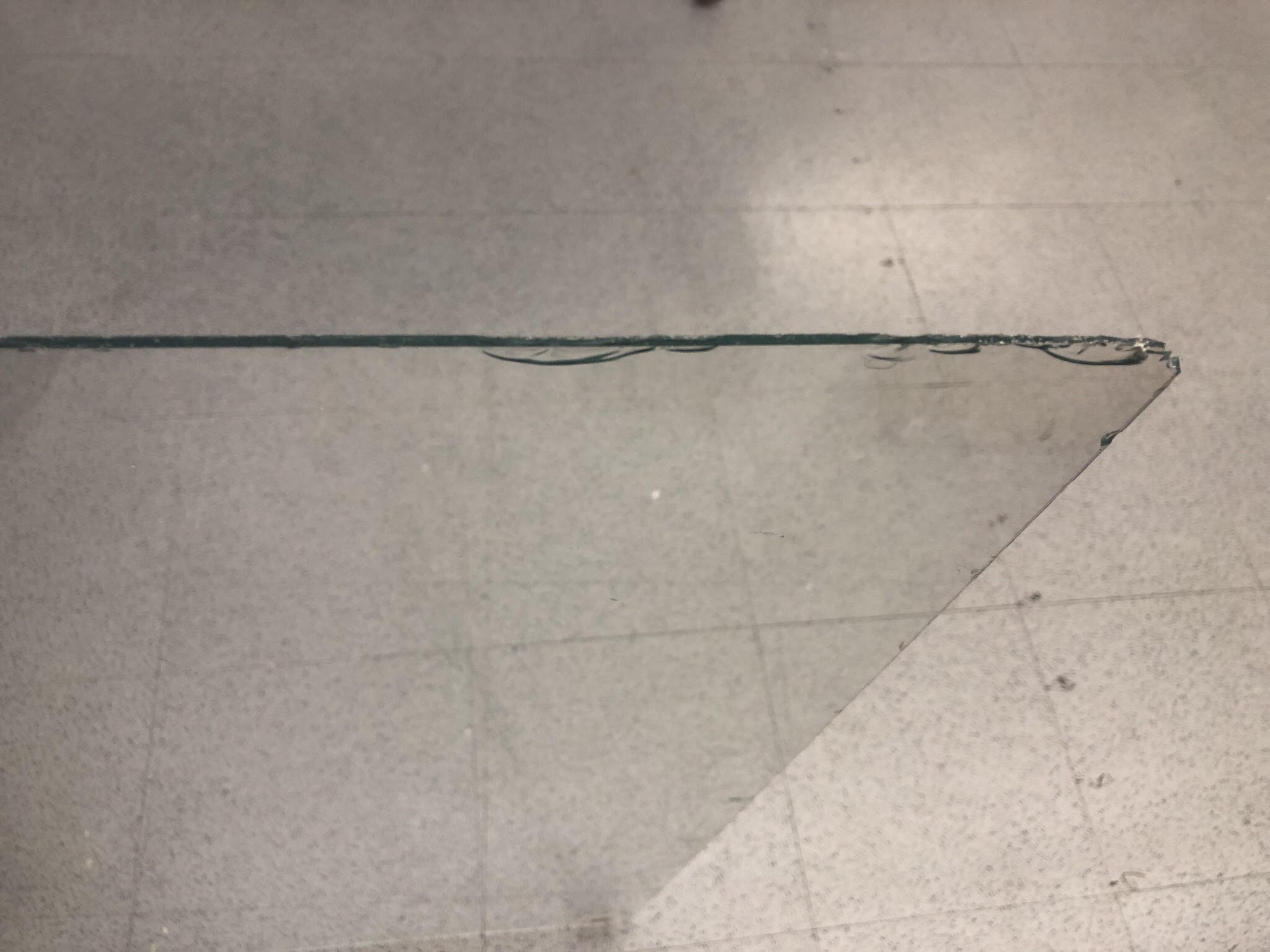}
	\caption{}
    \end{subfigure}
	\begin{subfigure}{0.24\textwidth}
	\centering
	\includegraphics[width=\textwidth]{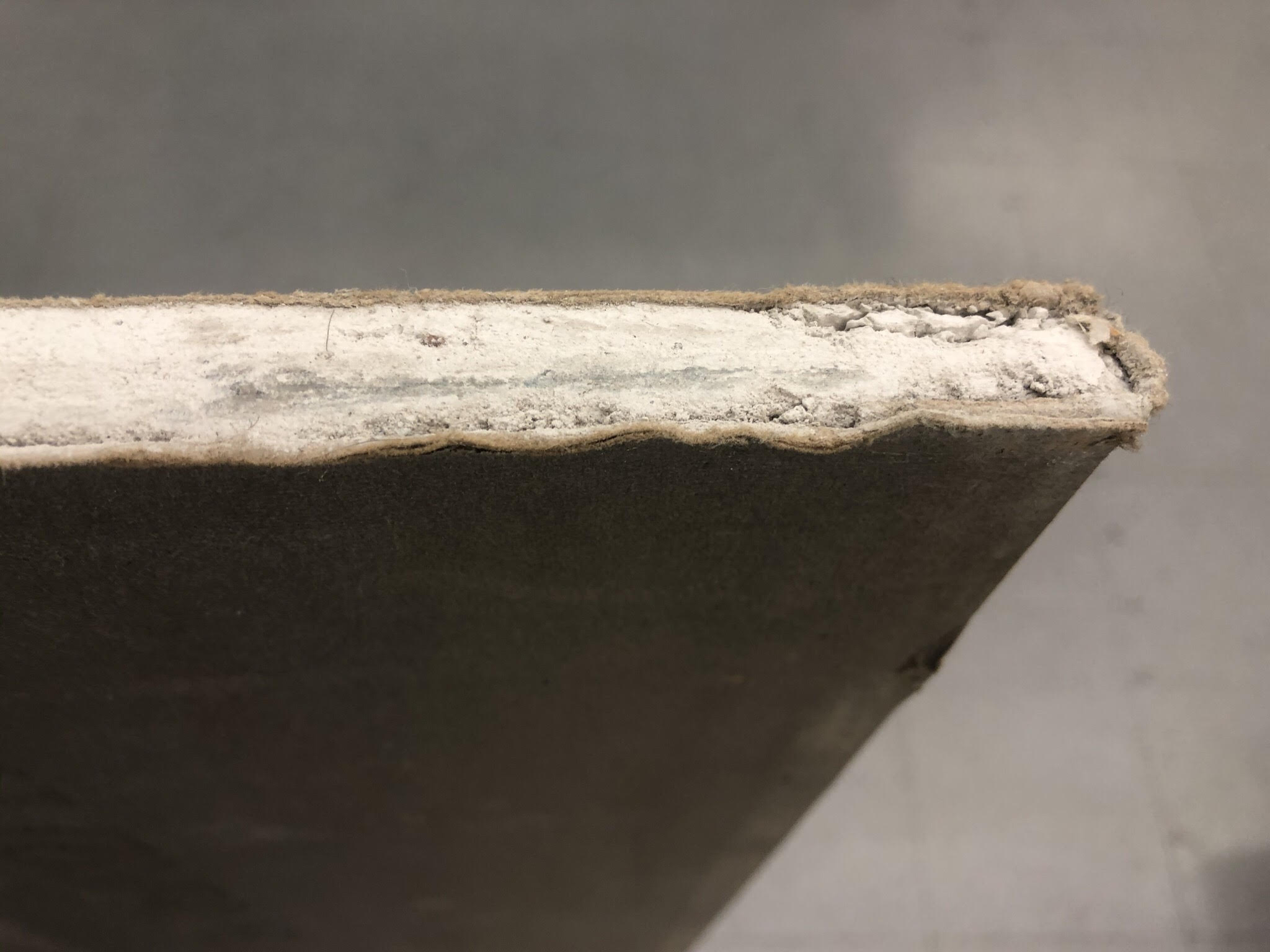}
	\caption{}
    \end{subfigure}
	\begin{subfigure}{0.24\textwidth}
	\centering
	\includegraphics[width=\textwidth]{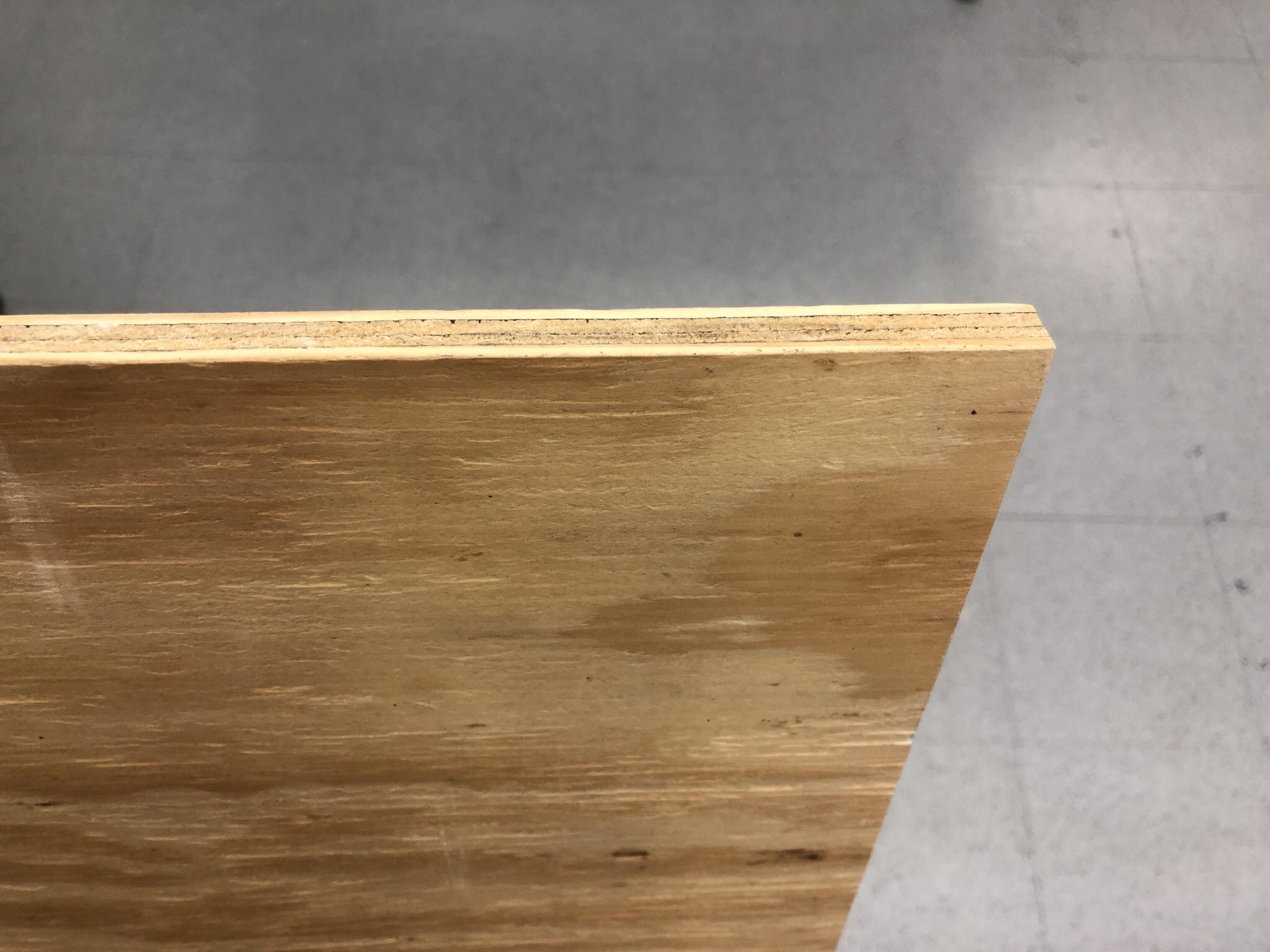}
	\caption{}
    \end{subfigure}
    \caption{The tested constructional materials, namely, (a) ceiling tile, (b) clear glass, (c) drywall, (d) plywood.\label{materials2}}
  \vspace{-3mm}
\end{figure}

\begin{figure*}[!t]
\centering
\includegraphics[width = .88\textwidth]{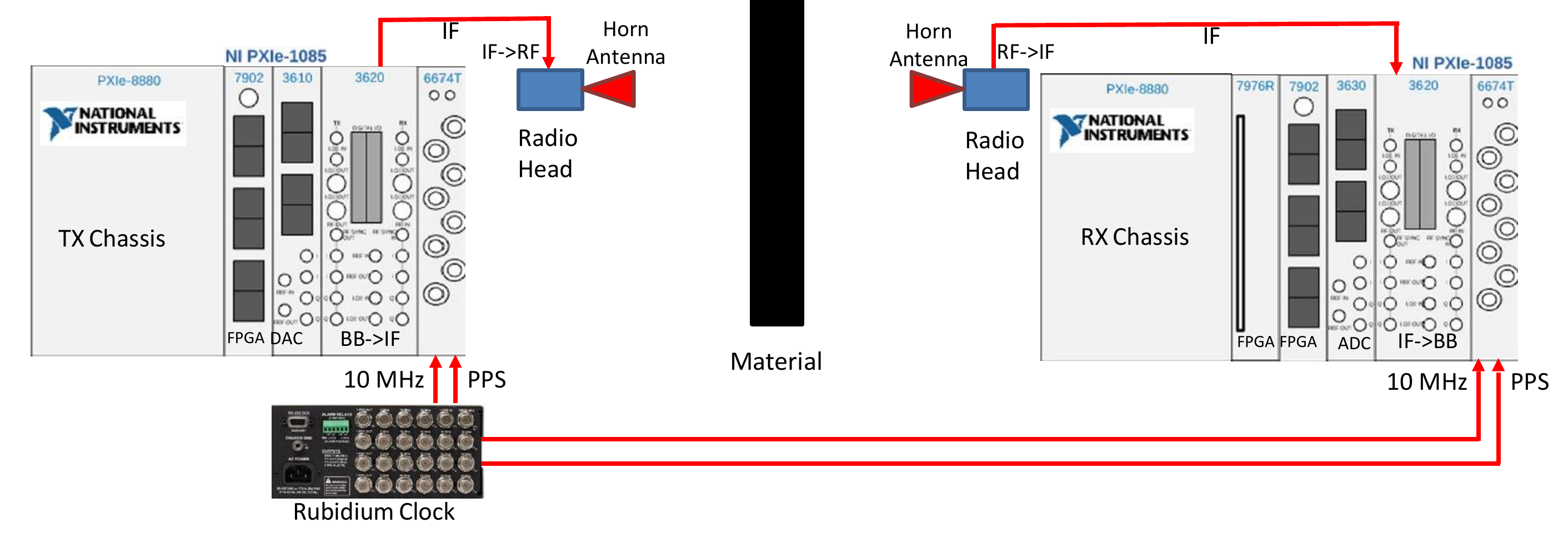}\vspace{-2mm}
\caption{Channel sounder block diagram for the NI PXI platform.} 
\label{Hardware_setup}\vspace{-4mm}
\end{figure*}

In this work, we focused on penetration loss measurements of several common constructional materials, as shown in Fig.~\ref{materials2}, including ceiling tile, drywall, clear glass, and plywood, at both mmWave band (28 and 39~GHz) and sub-THz  (120 and 144~GHz) bands. Measurements were carried out using a channel sounder based on NI's PXI  platform~\cite{NImmwave} (see Fig.~\ref{Hardware_setup}), with horn antennas on both Tx and Rx. We describe the signal generation and reception of our channel sounder in detail for each of the frequencies in consideration. 
Both penetration loss and attenuation were observed to increase with frequency, which were quantified for different materials. Results show that ceiling tile has the lowest penetration loss and attenuation at 28~GHz. At 144~GHz, plywood has the largest penetration loss of 16.068~dB, and clear glass has the highest attenuation of 27.633~dB/cm.

\section{Channel Sounder Setup and Configuration}

The block diagram of the channel sounder is given in Fig.~\ref{Hardware_setup}~\cite{erden2020correction}. It consists of NI PXIe-1085 Tx and Rx chassis, Tx and Rx radio heads, and a FS725 Rubidium~(Rb) clock~\cite{SRS}. 
The PXIe-1085 chasis performs a series of samplings and conversions as follows~\cite{erden2020correction}. The LabVIEW based sounder code periodically transmits a Zadoff-Chu (ZC) sequence of length 2048 that is processed by the root-raised-cosine (RRC) filter. The generated samples are uploaded to the PXIe-7902 FPGA and are sent to PXIe-3610 digital-to-analog (DAC) converter at a sampling rate of $f_{s}=3.072$~GS/s. 
The PXIe-3620 module up-converts the base-band signal to IF. 
For the 28~GHz and the 39~GHz setups, the baseband signal is multiplied with 3 times the first LO signal at $3.52$~GHz, to reach an IF of $10.56$~GHz. 
For the 120~GHz and the 144~GHz up-conversion, the IF is centered at 12~GHz, which is 3 times the LO signal at 4~GHz.
Tx radio head further up-converts the IF signal to RF. 

After sounding the channel, the Rx radio head down-converts the RF signal to IF, and the PXIe-3620 down-convert the IF to baseband. The PXIe-3630 analog-to-digital converter (ADC) samples the base-band analog signal with a sampling rate of $f_{s}=3.072$~GS/s. Our channel sounder thus provides $2/f_{s}=0.651$~ns delay resolution in the time domain. Therefore, any multipath component with a delay higher than $0.651$~ns can be resolved, which represents a path-length difference of $0.195$~m. The correlation and averaging are performed in PXIe-7902 FPGA, and the complex channel impulse response (CIR) samples are sent to the PXIe-8880 host PC for further processing and for saving to local disk.

The Rb clock is connected to PXIe-6674T modules at both the Tx and the Rx. The 10~MHz signal is used to generate the required local oscillator~(LO) signals, and the transmission at the Tx side and the reception at the RX side were triggered by the PPS signal. The calibration and the equalization are performed to eliminate the channel distortion due to the non-idealities of the hardware itself. Directional horn antennas are later connected to the mmWave radio heads at the Tx and the Rx sides. Time alignment is finally checked before the measurement to guarantee an accurate flight time of each peak in the channel impulse response.

We used different radio heads and horn antennas in penetration measurements at 28, 39, 120, and 144~GHz, as introduced in Section~\ref{28}, Section~\ref{39} and Section~\ref{144}, respectively.
The parameters of the horn antennas are provided in Table~\ref{antennas}. The base-band signals are the same for all of the four frequencies. The bandwidth of the RF signal centered at 28, 39, 120, and 144~GHz is 1.5~GHz regardless of the central frequency. The up-conversion and the down-conversion process of the radio heads for the corresponding four frequency bands are provided in Section~\ref{28}, Section~\ref{39} and Section~\ref{144}.

\begin{table}[b]
\footnotesize
\centering
\caption{Horn antenna parameters of 28, 39, and 120/144~GHz.} 
\label{antennas}
\begin{tabular}
{|p{1.1in}|p{0.55in}|p{0.55in}|p{0.6in}|} \hline 
 \textbf{Operating Frequency} & \textbf{28~GHz} & \textbf{39~GHz} & \textbf{120/144~GHz}\\ \hline
Model  &  SAR-1725-34KF-E2  &  SAR-2013-222F-E2  &  VDI WR-6.5\\ \hline 
Antenna gain (dBi)  &  17  &  20  &  21 \\ \hline 
3 dB beam width in E plane (degree)  &  26  &  15  &  13 \\ \hline 
3 dB beam width in H plane (degree)  &  24  &  16  &  13 \\ \hline 
Antenna maximum linear dimension (cm)  &  4.072  &  4.384  &  1.080 \\ \hline 
Antenna far-field (cm) &  30.973  &  50.005 &  11.205\\ \hline 
\end{tabular}
\end{table}

\subsection{28~GHz Tx/Rx Setup}
\label{28}

At 28~GHz, NI mmRH-3642 was the Tx radio head and NI mmRH-3652 was the Rx radio head. The Tx radio head up-converts the IF signal by multiplying 8 times of the second LO at $4.82$ GHz. The lower side-band is centered at 28~GHz and the upper side-band is filtered out in the radio head. At the Rx, the Rx radio head down-converts the RF signal to IF. SAR-1725-34KF-E2 horn antennas are connected to the Tx and the Rx radio head with 17~dBi gain, 26 degree and 24 degree half power beam-widths in the elevation and the azimuth planes, and a maximum linear dimension of $4.072$~cm.

\subsection{39~GHz Tx/Rx Setup}
\label{39}

At 39~GHz, NI mmRH-3643 is the Tx radio head and NI mmRH-3653 is the Rx radio head. The Tx radio head up-converts the IF signal by multiplying 6 times of the second LO at $4.456$ GHz. The upper side-band is centered at 39~GHz and the lower side-band is filtered out in the radio head. At the Rx, the Rx radio head down-converts the RF signal to IF. SAR-2013-222F-E2 horn antennas shown in Table~\ref{antennas} are connected to the Tx and the Rx radio head with 20~dBi gain, 15 degree and 16 degree half power beam-width in the elevation and azimuth planes, and a maximum linear dimension of $4.384$~cm.

\begin{figure*}[!b]\vspace{-4mm}
\centering
\includegraphics[width = .91\textwidth]{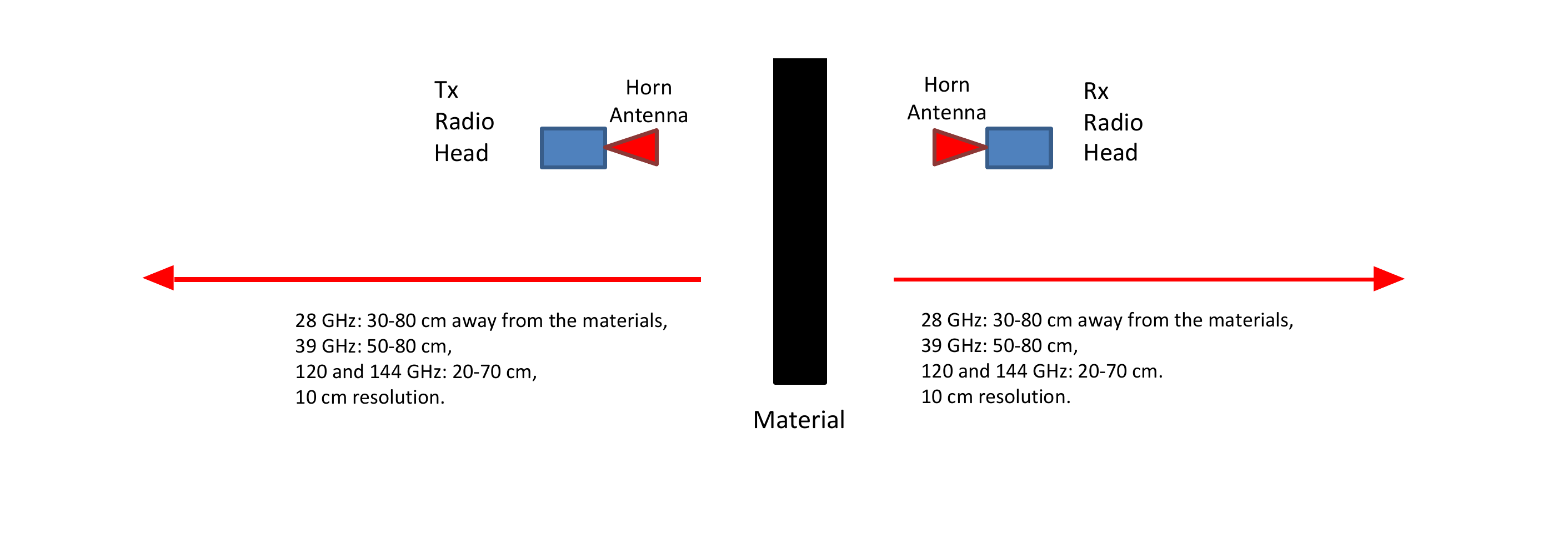}\vspace{-8mm}
\caption{Penetration loss measurement setup, with the construction material located between the Tx and the Rx antennas.} 
\label{penetration_setup}\vspace{-3mm}
\end{figure*}

\subsection{120 GHz and 144~GHz Tx/Rx Setup}
\label{144}

At 120~GHz and 144~GHz, VDI WR6.5CCU is used as the Tx radio head and VDI WR6.5CCD is used as the Rx radio head. The Tx radio head up-converts the IF signal by multiplying 12 times of the second LO at 9~GHz for 120~GHz, and at 11~GHz for 144~GHz. The upper side-band is centered at 144~GHz and the lower side-band is filtered out by WR6.5BPFE140-148 bandpass filter, located between the Tx radio head and the Tx antenna. For the 120~GHz signal, the lower side-band is filtered out by WR6.5BPFE116-123 bandpass filter. At the Rx, the Rx radio head down-converts the RF signal to IF. The VDI WR-6.5 horn antennas are connected to the Tx and the Rx radio head with 21~dBi gain, as shown in Table~\ref{antennas}, with 13 degree half power beam-widths in both elevation and azimuth planes, and a maximum linear dimension of 1.08~cm.

\section{Penetration Loss Measurement Procedure}

\begin{table}[b]
\footnotesize
\renewcommand{\arraystretch}{1.2}
\centering
\caption{List of materials used in the measurement~\cite{hosseini2020attenuation}.} 
\label{materials}
\begin{tabular}
{|p{0.4in}|p{1.25in}|p{0.55in}|p{0.5in}|} \hline 
\textbf{Material} & \textbf{Model} & \textbf{Dimension} & \textbf{Thickness}\\ \hline
Ceiling tile & Armstrong Ceilings Acoustic Panel Ceiling Tiles & $121.9$~cm $\times$ $60.9$~cm & $1.13$~cm \\ \hline 
Clear glass & Gardner Glass Products Clear Glass & $76.2$~cm $\times$ $91.4$~cm & $0.18$~cm \\ \hline 
Drywall & ToughRock Fireguard Drywall Panel & $121.9$~cm $\times$ $243.8$~cm & $1.31$~cm \\ \hline 
Plywood (4-ply) & Plytanium $15/32$ CAT PS1-09 Square Structural Pine Sheathing & $121.9$~cm $\times$ $243.8$~cm & $1.12$~cm \\ \hline 
\end{tabular}
\end{table}

The penetration loss measurements of common constructional materials were conducted in an indoor environment in Engineering Building 2 at NC State University, Raleigh, NC. List of the materials used for the penetration loss measurements is given in Table~\ref{materials}, along with their dimensions and models. Fig.~\ref{penetration_setup} illustrates the measurement scenario. The Tx and the Rx were first aligned to each other, and the LOS received power without blockage was measured. Then, the material was introduced in the center of the Tx and the Rx, and was further aligned to both Tx and Rx so that Tx and Rx were pointing directly to the material at 0~degree incident angle. The NLOS received power with the blockage was then measured. After all the received power was measured with different materials' blockage, both the Tx and the Rx were moved to the next position and we repeated the measurement procedure mentioned above.

\begin{figure}[!t]
\centering
\includegraphics[width = .44\textwidth]{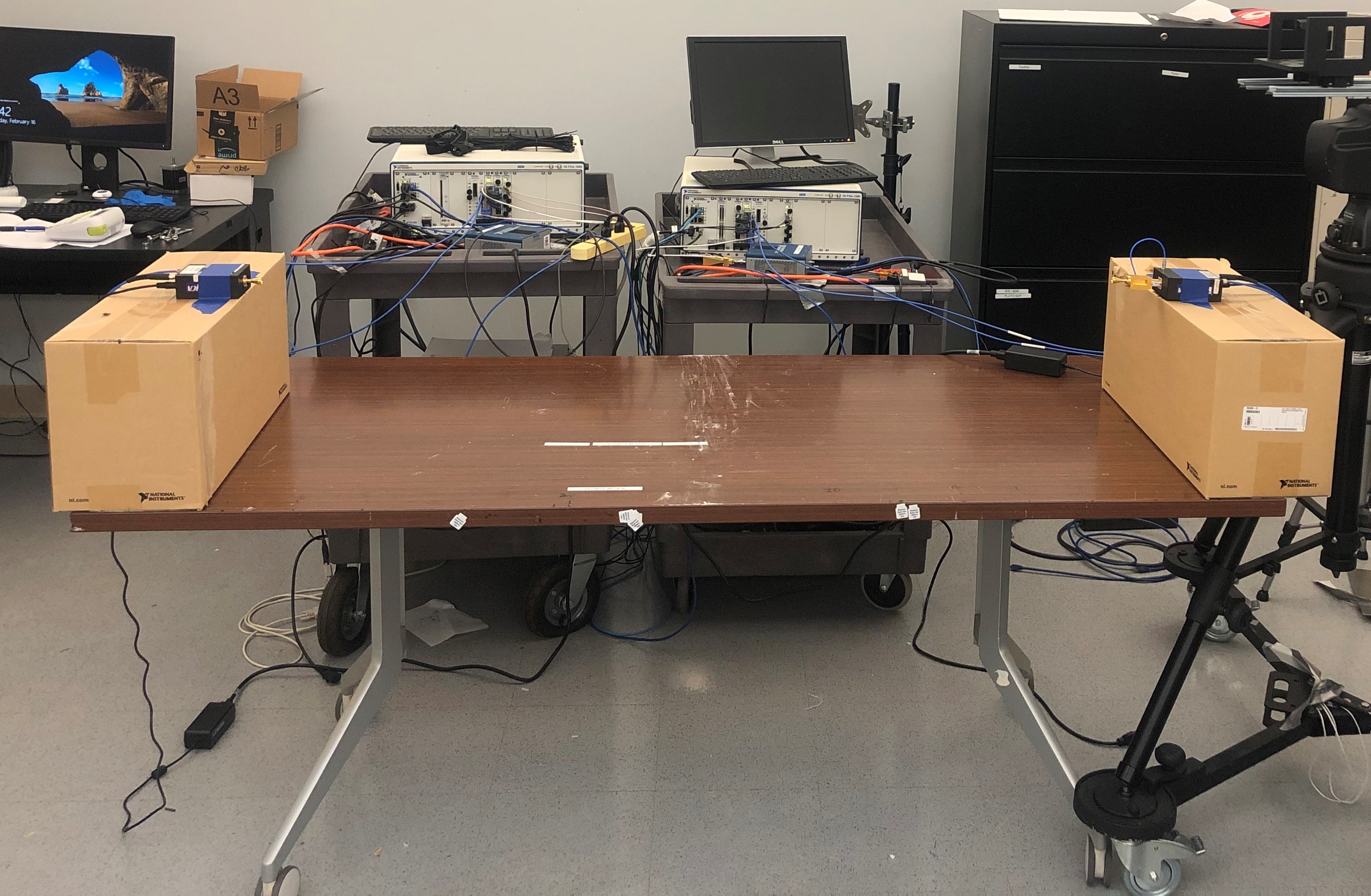}
\caption{120~GHz and 144~GHz measurements setup.} 
\label{desk}\vspace{-5mm}
\end{figure}

 
For the 28~GHz and the 39~GHz measurements, a maximum distance of 80~cm from the Tx and the Rx to the material was chosen, thus the cross-section of the material and the antenna main lobe would be much smaller than the dimensions of the material~\cite{hosseini2020attenuation}, and the effect of diffraction and the edge scattering was minimized. For the 120~GHz and 144~GHz measurements, the maximum distance was chosen as 70~cm due to the length limit of the desk that held the materials and the radio heads, as shown in Fig.~\ref{desk}.
The minimal distance to the materials was chosen according to the far field region $R$ of the horn antennas, which is calculated as:
\begin{align}
\mathrm{R}> \frac{2D^2}{\lambda}~,\quad  
\mathrm{R} \gg D~,\quad   
\mathrm{R} \gg \lambda~,  
\end{align}
where $\lambda$ is the wave length of the radiated signal and $D$ is the maximum linear dimension of the antenna. We make sure that the materials are in the far field region of both the Tx and the Rx antenna because in this region radiated fields dominate, and the radiation pattern does not change shape with distance.

The calculated far fields for the three different antennas are shown in Table~\ref{antennas}. The 28~GHz antennas have far field region greater than 30~cm. Therefore, the received power with and without blockage was measured when the material is 30~cm to 80~cm away from both the Tx and the Rx, with a resolution of 10~cm. The 39~GHz antennas have far field region greater than 50~cm, and therefore we measured the received power with and without blockage when the material is 50~cm to 80~cm away from both the Tx and the Rx with 10~cm resolution. The far field region of the 120~GHz and the 144~GHz antennas are greater than 11~cm, and hence, the received power with and without blockage was measured when the material is 20 to 70~cm away from the Tx and the Rx, with 10~cm resolution. Transmit power was set to -10~dBm for all the measurements to protect the receiver at close distances.

The received power with blockage $P_{r,{\textrm{LOS}}}$ and without blockage $P_{r,{\textrm{blocked}}}$ can be calculated as:
\begin{align}
\mathrm P_{r,{\textrm{LOS}}}~({\rm dB})  &=  P_{t} + G_{t} + G_{r} - \rm FSPL~,  \nonumber 
\\
\mathrm P_{r,{\textrm{blocked}}}~({\rm dB})  &=  P_{t} + G_{t} + G_{r} - \rm FSPL  - \mathrm{Penetration~Loss}, \nonumber 
\end{align} 
where $P_{t}$ is the transmitted power, $G_{t}$ and $G_{r}$ are the Tx and the Rx antenna gain, FSPL is the free space path loss which is the same for both cases since the signals travel the same distances.

Therefore, the penetration loss can be further derived from the measured LOS power without blockage $P_{r,{\textrm{LOS}}}$ and the measured blocked LOS received power $P_{r,{\textrm{blocked}}}$ at the same Tx and Rx separation, using:
\begin{equation}
\mathrm{Penetration~Loss~(\rm dB)} = P_{r,{\textrm{LOS}}} - P_{r,{\textrm{blocked}}}~.  \label{GrindEQ__6_}
\end{equation} 

\begin{figure}
	\centering
	\begin{subfigure}{0.4\textwidth}
	\centering
	\includegraphics[width=\textwidth]{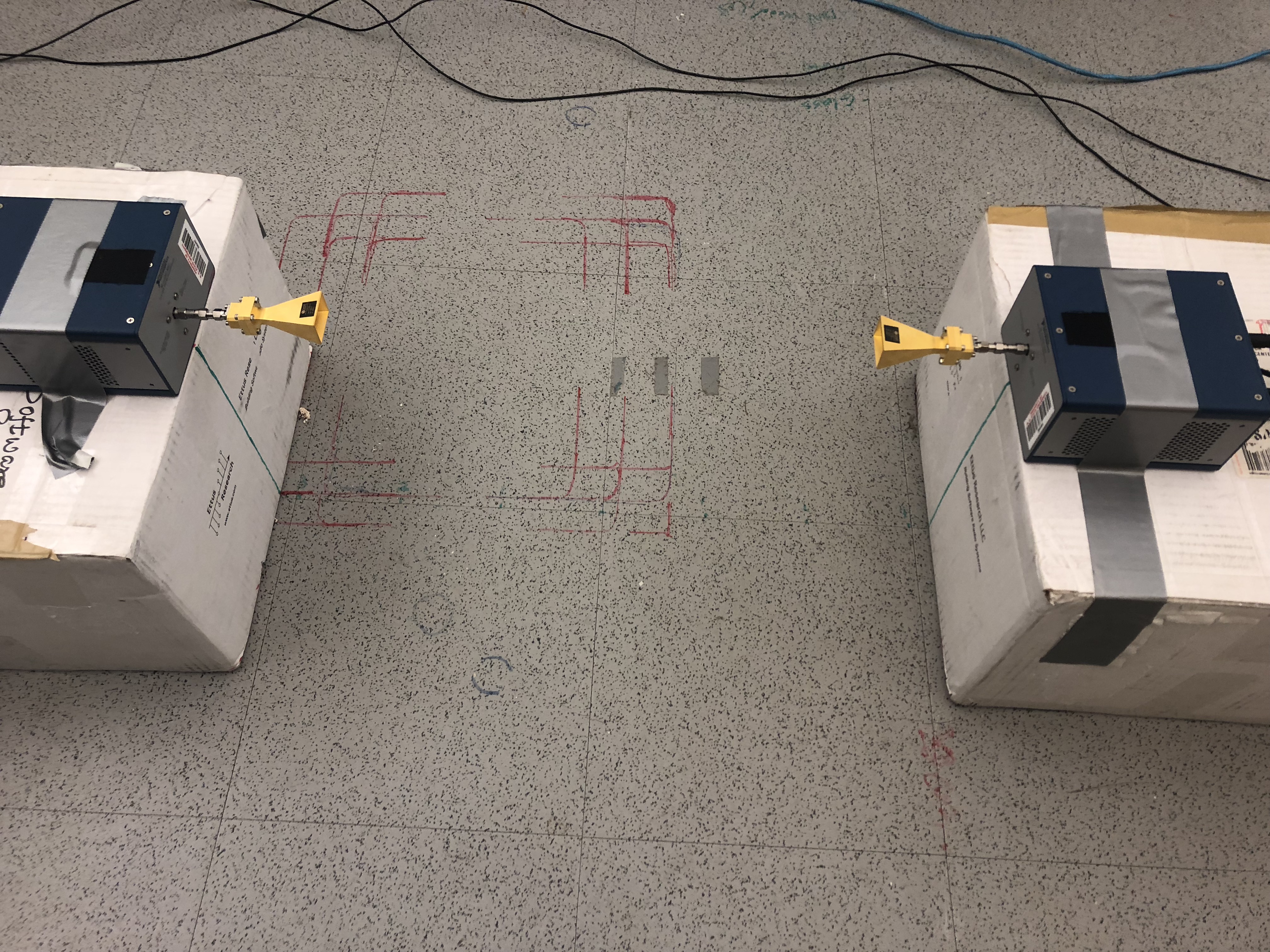}
	\caption{}
	\vspace{1mm}
    \end{subfigure}
    \begin{subfigure}{0.4\textwidth}
	\centering
	\includegraphics[width=\textwidth]{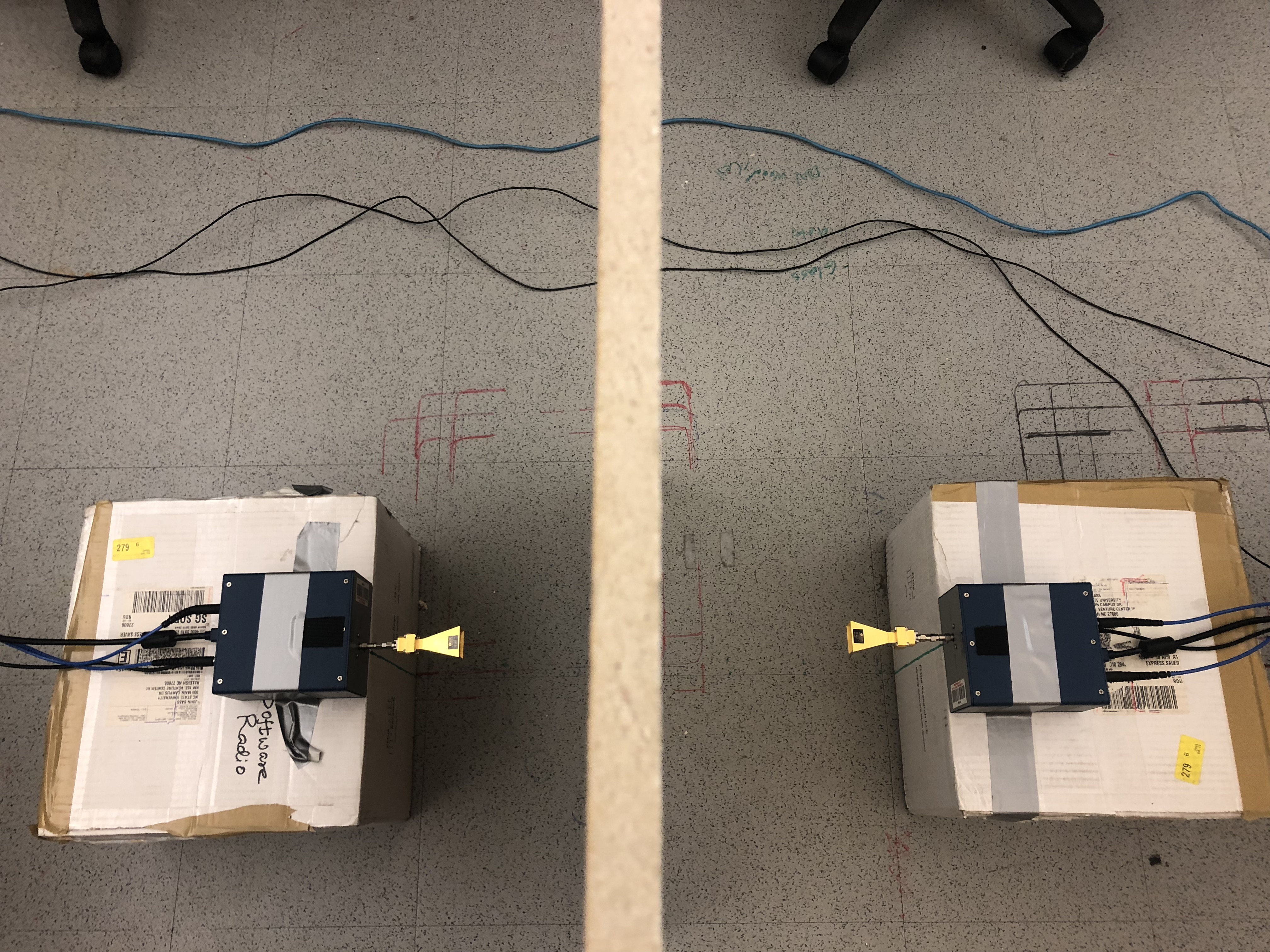}
	\caption{}
	\vspace{1mm}
    \end{subfigure}
    \caption{Measurement (a) without blockage, (b) with material blockage.}
    \label{measurement}
  \vspace{-3mm}
\end{figure}

Fig.~\ref{measurement}(a) and Fig.~\ref{measurement}(b) illustrate an example measurement setup for the LOS case and the blocked LOS case, respectively, for the 144~GHz measurements. Note that the Tx and the Rx radio heads were placed above the ground high enough so that in the unblocked scenario the ground reflection is resolvable to the LOS component at close measured distance. At larger distance, the ground reflection is not resolvable to the LOS component, but its effect is minimized by the narrow beam of our antennas. Received power was directly read from the first arriving multipath component in the CIR. To get a more static and accurate result, the CIR in each measurement was averaged over 32 scans by the sounder code.


\section{Measurement Results and Analysis}

\begin{table*}[b]
\footnotesize
\renewcommand{\arraystretch}{1.1}
\centering
\caption{Measurement results of average received power, penetration loss, standard deviation of penetration loss, and attenuation at 28~GHz, 39~GHz, 120~GHz, and 140~GHz for ceiling tile, clear glass, drywall, and plywood.} 
\label{result}
\begin{tabular}
{|c|c|c|c|c|c|c|} \hline
 \textbf{Material} & \textbf{Frequency} & \textbf{$P_{r,{\textrm{LOS}}}$~(dBm)} & \textbf{$P_{r,{\textrm{blocked}}}$~(dBm)} & \textbf{Penetration Loss~(dB)} & \textbf{Standard Deviation~(dB)} & \textbf{Attenuation~(dB/cm)}\\ \hline
\multirow{4} {*} {Ceiling tile} & 28~GHz & $-37.722$ & $-38.123$ & $0.401$ & $0.027$ & $0.355$ \\ \cline{2-7}
& 39~GHz & $-39.005$ & $-39.853$ & $0.848$ & $0.083$ & $0.75$ \\ \cline{2-7}
& 120~GHz & $-38.279$ & $-40.993$ & $2.714$ & $0.269$ & $2.402$ \\ \cline{2-7}
& 144~GHz & $-39.405$ & $-43.524$ & $4.119$ & $0.925$ & $3.645$ \\ \hline
\multirow{4} {*} {Clear glass} & 28~GHz & $-37.722$ & $-38.513$ & $0.791$ & $0.009$ & $4.394$ \\ \cline{2-7}
& 39~GHz & $-39.005$ & $-42.777$ & $3.772$ & $0.087$ & $20.956$ \\ \cline{2-7}
& 120~GHz & $-38.279$ & $-43.007$ & $4.728$ & $0.083$ & $26.267$ \\ \cline{2-7}
& 144~GHz & $-39.405$ & $-44.379$ & $4.974$ & $0.786$ & $27.633$ \\ \hline
\multirow{4} {*} {Drywall} & 28~GHz & $-37.722$ & $-38.858$ & $1.137$ & $0.037$ & $0.868$ \\ \cline{2-7}
& 39~GHz & $-39.005$ & $-40.503$ & $1.498$ & $0.103$ & $1.144$ \\ \cline{2-7}
& 120~GHz & $-38.279$ & $-41.196$ & $2.917$ & $1.641$ & $2.227$ \\ \cline{2-7}
& 144~GHz & $-39.405$ & $-43.193$ & $3.788$ & $2.789$ & $2.892$ \\ \hline
\multirow{4} {*} {Plywood} & 28~GHz & $-37.722$ & $-40.838$ & $3.116$ & $0.934$ & $2.782$ \\ \cline{2-7}
& 39~GHz & -$39.005$ & $-43.274$ & $4.269$ & $0.597$ & $3.812$ \\ \cline{2-7}
& 120~GHz & $-38.279$ & $-52.616$ & $14.337$ & $4.276$ & $12.801$ \\ \cline{2-7}
& 144~GHz & $-39.405$ & $-55.473$ & $16.068$ & $3.845$ & $14.346$ \\ \hline
\end{tabular}\vspace{-3mm}
\end{table*}

The measurement results were averaged over all the measured positions. The averaged received power, penetration loss, standard deviation for the penetration loss, and attenuation for different materials at 28~GHz, 39~GHz, 120~GHz, and 144~GHz are provided in Table~\ref{result}.
The averaged received power of the LOS path (at the same travel distance as the blocked case) of all the measured positions is denoted by $P_{r,{\textrm{LOS}}}$. On the other hand, $P_{r,{\textrm{blocked}}}$ is the averaged received power of the NLOS ray that is blocked by the materials. Penetration loss was calculated from \eqref{GrindEQ__6_} based on the difference of the received power with and without blockage. The standard deviation indicated the variation of the calculated penetration loss. The attenuation in~dB/cm, which represents the penetration loss of unit thickness of the material, is the averaged penetration loss over material thickness.

As shown in Table~\ref{result}, the lowest penetration loss was measured as $0.401$~dB for ceiling tile at 28~GHz and the highest was $16.068$~dB at 144~GHz. At all four frequencies, ceiling tile has the lowest penetration loss ranging from $0.401$~dB at 28~GHz to $4.119$~dB at 144~GHz among all the materials, while plywood has the largest penetration loss ranging from $3.116$~dB at 28~GHz to $16.068$~dB at 144~GHz. On the other hand, the attenuation of clear glass is the highest compared with other materials at each frequencies. A largest attenuation of $27.633$~dB/cm was observed at 144~GHz for clear glass. At 28~GHz and 39~GHz, the standard deviation of the penetration loss for all the materials was all less than 1~dB. However, for 120 and 144~GHz the standard deviation increased, with a maximum of $4.276$~dB for plywood at 120~GHz. Plywood has the largest standard deviation at all frequencies, which could somehow be explained by the diffuse reflection at the rough surface of the plywood. 
Besides, as wavelength gets smaller, the surface of the material is rougher for the incident ray, thus causing an increased variation in penetration loss at higher frequencies.

\begin{figure*}
	\centering
	\begin{subfigure}{0.42\textwidth}
	\centering
	\includegraphics[width=\textwidth]{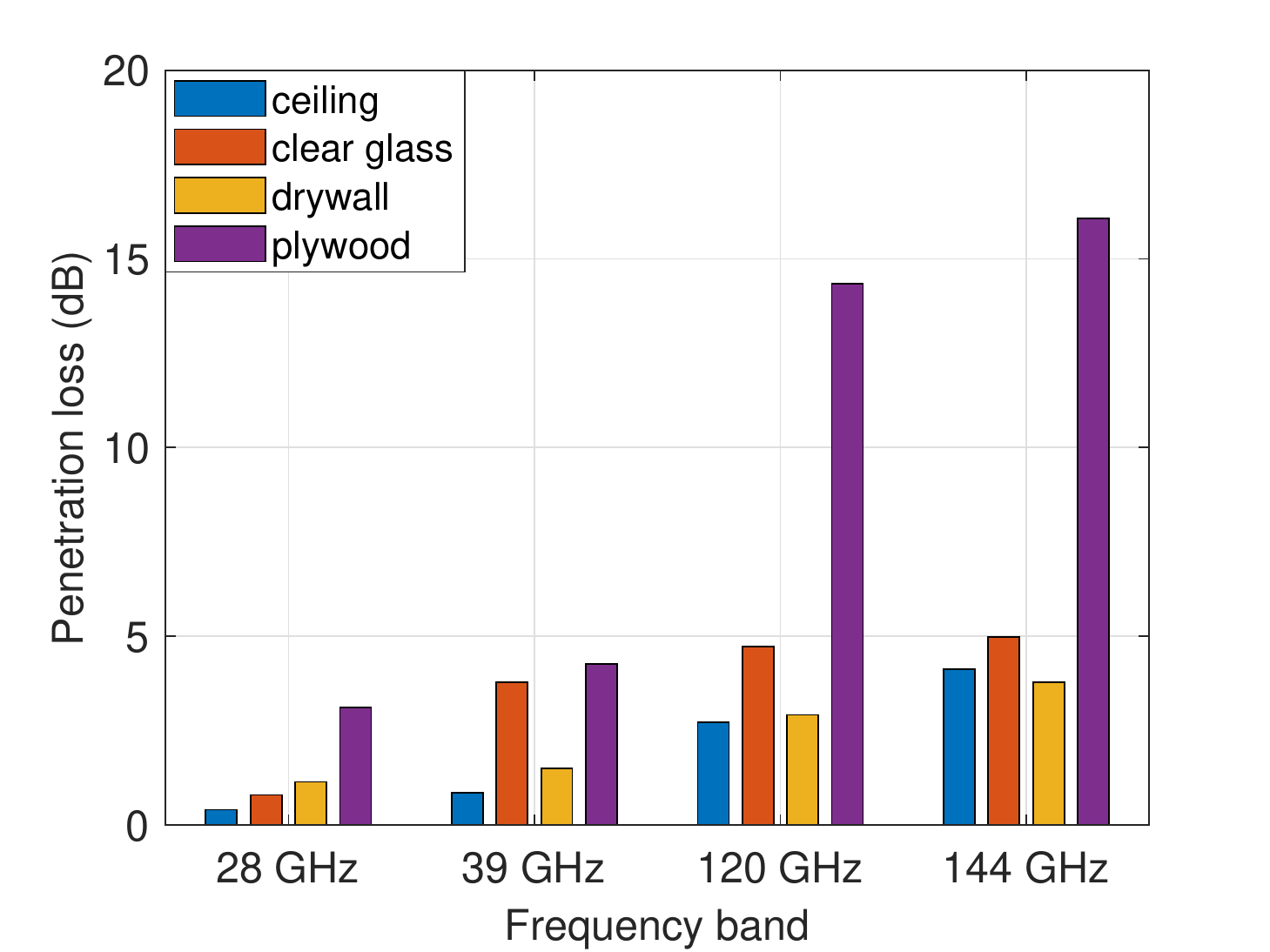}
	\caption{}
   \end{subfigure}
    \begin{subfigure}{0.42\textwidth}
	\centering
	\includegraphics[width=\textwidth]{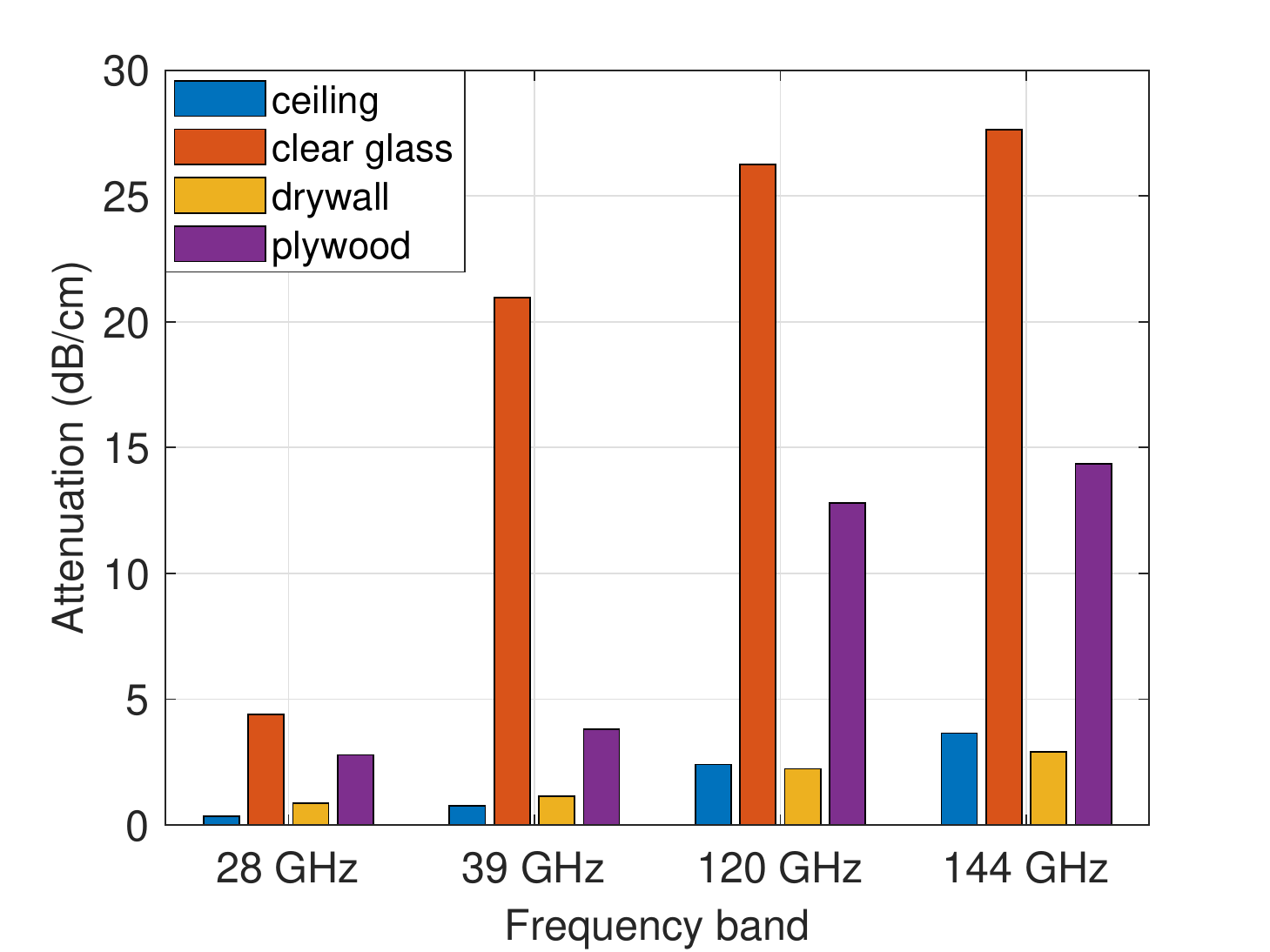}
	\caption{}
    \end{subfigure}
    \caption{The averaged results of (a) penetration loss, (b) attenuation for ceiling tile, clear glass, drywall, and plywood at 28, 39, 120, and 144~GHz.}
    \label{result_all}
  \vspace{-4mm}
\end{figure*}

The averaged penetration loss and attenuation of different materials at four different frequencies were plotted in Fig.~\ref{result_all}.
The overall results show that the penetration loss and the attenuation increases with frequency. Fig.~\ref{result_all}(a) indicates that plywood has an attenuation larger than 14~dB, which would seriously degrade the performance of 120~GHz and 144~GHz  NLOS signal propagation when penetrating through walls and floors, while all the other materials have attenuation less than 5~dB at all frequencies. 
Fig.~\ref{result_all}(b) shows the effect of material thickness. Although the penetration loss of clear glass board was less than 5~dB, the attenuation was larger than 20~dB/cm at 39~GHz, 120~GHz, and 144~GHz. A thin glass window or wall is therefore necessary for better outdoor to indoor coverage at these three bands.


\section{Conclusion}

In this work, we measured the penetration loss of several common constructional materials at 28~GHz, 39~GHz, 120~GHz, and 144~GHz. The maximum distance from the Tx and the Rx to the materials was chosen to eliminate the edge scattering effect. The minimal distance from the Tx and the Rx to the materials was calculated to ensure both the Tx and the Rx antennas were in the far field region. Ground reflection was minimized by using narrow beam antennas. The measured constructional materials were ceiling tile, glass, drywall, and plywood. The penetration loss results, after being averaged over all the measured positions, ranged from $0.401$~dB for ceiling tile at 28~GHz to $16.068$~dB for plywood at 144~GHz. The highest attenuation of $27.633$~dB/cm was observed for clear glass at 144~GHz. Overall, the penetration loss and attenuation increases with frequency for all the materials. These results could be helpful for developing mmWave and THz channel models that accurately estimate the penetration loss of common building materials. They can further contribute to link budget calculation for future 5G/6G deployments in indoor environments and for developing accurate penetration models for ray-tracing simulations.

\section*{Acknowledgement}

This work was supported in part by NASA under the Award ID NNX17AJ94A, and in part by DOCOMO Innovations, Inc.

\bibliographystyle{IEEEtran}
\bibliography{IEEEabrv,reference}

\end{document}